\documentclass[review]{elsarticle}

\makeatletter
\def\ps@pprintTitle{%
 \let\@oddhead\@empty
 \let\@evenhead\@empty
 \def\@oddfoot{\centerline{\thepage}}%
 \let\@evenfoot\@oddfoot}
\makeatother

\usepackage{times}
\usepackage{epsfig}
\usepackage{graphicx}
\usepackage{amsmath}
\usepackage{amssymb}
\usepackage{xcolor}
\usepackage{filecontents}

\usepackage[linesnumbered,vlined,ruled,norelsize]{algorithm2e}
\usepackage{hyperref}

\journal{Vision Research}

\DeclareMathOperator*{\argmin}{arg\,min}

\def\R{\mathbb{R}}
\def\tp{^{\mathsf{T}}}

\newcommand{\mv}[1]{\mathbf{#1}}

\biboptions{authoryear}

\begin{document}

\begin{frontmatter}

\title{Maps of Visual Importance}

\author[mymainaddress]{Xi Wang\corref{mycorrespondingauthor}}
\cortext[mycorrespondingauthor]{Corresponding author at Computer Graphics, Technische Universit{\"{a}}t Berlin, Germany.}
\ead{xi.wang@tu-berlin.de}

\author[mymainaddress]{Marc Alexa}
\address[mymainaddress]{Technische Universit{\"{a}}t Berlin, Germany}

\begin{abstract}
The importance of an element in a visual stimulus is commonly associated with the fixations during a free-viewing task. We argue that fixations are not always correlated with attention or awareness of visual objects. 
We suggest to filter the fixations recorded during exploration of the image based on the fixations recorded during recalling the image against a neutral background. This idea exploits that eye movements are a spatial index into the memory of a visual stimulus.
We perform an experiment in which we record the eye movements of 30 observers during the presentation and recollection of 100 images. The locations of fixations during recall are only qualitatively related to the fixations during exploration. We develop a deformation mapping technique to align the fixations from recall with the fixation during exploration. This allows filtering the fixations based on proximity and a threshold on proximity provides a convenient slider to control the amount of filtering. Analyzing the spatial histograms resulting from the filtering procedure as well as the set of removed fixations shows that certain types of scene elements, which could be considered irrelevant, are removed. In this sense, they provide a measure of importance of visual elements for human observers. 
\end{abstract}


\end{frontmatter}

\section{Introduction}
\label{sec:intro}

The relative importance of visual elements presented on a display is useful in a variety of applications. 
In the fields of visualization and human-computer interaction, it is often a design goal to make specific elements clearly visible. Consider for example the visualization of medical imaging data: it is important that anomalies are recognized by the physician inspecting the data. Likewise, in computer graphics or image and video coding, it is of interest what visual objects are important or unimportant in order to allocate resources such as computation time (for example, for rendering 3D objects in an augmented reality scenario) or bandwidth (as in video coding).

Note that in these applications we are not just interested in whether visual elements have been processed by an observer, but we need a map of the relative importance of regions of the stimulus. The common approach for creating such maps is based on eye tracking and accumulating fixations from several observers~\citep{Bruce2005,judd2009,judd2012,Coutrot2014,xu2014,borji2015}
The underlying reasoning is that most visual information is processed during fixating~\citep{matin1974}, and saccade target selection is likely optimized to attend to the most important visual elements first~\citep{itti2000,Parkhurst2002}. A variety of factors have been shown to affect the selection~\citep{Krieger00,einhaeuser2008,kienzle2009,tatler2011,schomaker2017}.

We are nonetheless unconvinced that fixations alone are sufficient for creating maps of importance of visual elements, for several reasons: fixated locations are not necessarily attended; and the frequency and duration of fixations is not always correlated with the importance of an object as assigned by an observer. Here, we suggest a new experimental setup that creates a map based on whether visual elements have been encoded and deemed important. 

\subsection{Fixations vs.\ visual attention}

It is well known that observers often fail to detect seemingly prominent visual objects, typically because they are given a task for which the object is irrelevant -- an effect called \emph{inattentional blindness}~\citep{SIMONS2000}. One may speculate that the specific task leads to a lack of fixations on the unexpected object, yet this is not the case. In particular, \cite{MEMMERT2006} repeated the famous Gorilla experiment while tracking the gaze of the observers. They found that observers fixate the Gorilla a significant amount of time and that recognizing the Gorilla was independent of the fixations. \cite{KOIVISTO2004} used a more controlled setup with a screen presentation of the stimulus. Their particular findings are that \emph{visual attention} and fixations were uncorrelated under the specific experimental conditions (they note, however, that this might not be common under natural viewing conditions).

\subsection{Fixation statistics vs.\ importance}

A variety of factors influence the statistics of fixations, some of which may be uncorrelated with the importance of a scene element that is later attributed by an observer. 

To begin with, a confounding factor is the experimental setup for gathering fixations: it is common to only consider fixations in the first few seconds after the onset of the stimuli. This is based on the intuitive idea that the most important elements would attract visual attention first. Indeed, it has been shown that fixation locations relate less to image statistics over time~\citep{itti2000,Parkhurst2002}. This approach, on the other hand, limits the number of locations extracted, as each fixation has a certain minimal duration (typically not less than 150ms~\citep{Pannasch08}). Also the distance between fixations (the saccade amplitude) is limited to about $7^\circ$ visual angle. 
These characteristics of fixations, saccades, and the time in which fixations are collected have an important consequence: only few elements can be visited and, since the distribution of fixations is constrained but not the distribution of scene elements, it is not clear that all fixations indeed correspond to visually important elements.
The common strategy for coping with this potential problem is to average over the data of repeated trials and several observers~\citep[e.g.][]{vanderLinde2009,Ramanathan2010,koehler2014}. It remains unclear why averaging would result in a measure of importance. 

\begin{figure}[t]
\begin{center}
   \includegraphics[width=0.6\linewidth]{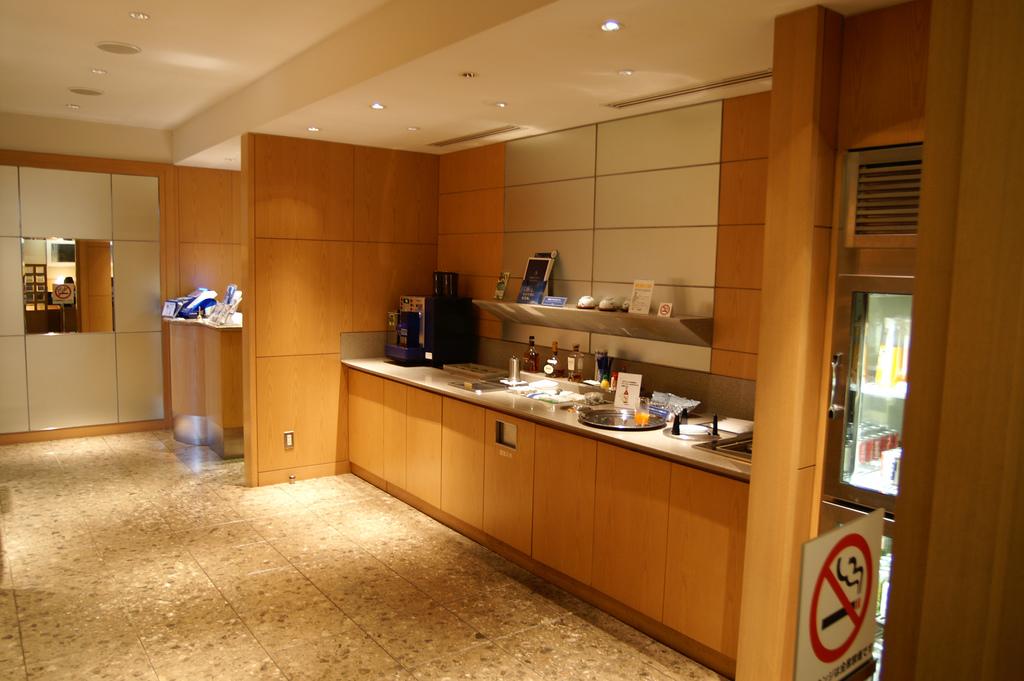}
\end{center}
   \caption{An example of mirror scene.}
\label{fig:mirror}
\end{figure}

More importantly, it is known that fixations may extend or accumulate for a variety of reasons, such as generally being biased towards the center of the field of view~\citep{Tatler:2007}, being longer for scene elements that are unlikely or improbable given the scene context~\citep{Loftus:1978,DeGraef1990}, or following the gaze of humans depicted in the scene~\citep{Ricciardelli2002,Hutton2011}. We also speculate that visual elements that are complicated (e.g., cluttered or  obstructed) require longer or more fixations. In all of these cases, the importance of the visual element depends on a variety of factors, and assigning high importance based on just the number or duration of fixations is unwarranted. As an example for this statement consider a mirror in the scene (see Figure ~\ref{fig:mirror}): the mirror likely draws fixations because it reflects a variety of interesting objects. We would suspect, once the region is being decoded as a mirror it likely decreases in importance, either because the objects are visible in the scene and can be attended to directly, or because understanding the scene through a mirror is too complicated to be attempted during a brief exposure to the stimulus.  

\subsection{Importance of fixation based on recall}

In this work, we distinguish among fixated regions that do and do not make it into the visual memory; and among the ones being stored we create a map that reflects their relative importance. 
The contribution of the paper is an experimental setup and an algorithm for post-processing that leads to such maps. 

The idea is to base importance on the motion of the eyes during the \emph{recall} of the image. It is know that the eyes move during recalling an image. In the absence of other visual features (i.e.\ while looking at nothing), the motion of the eyes is related to the motion of the eyes while the stimulus was present. In particular, fixations during recall are related to fixations during exposure with the stimulus~\citep{johansson2014, laeng2014}. We want to use fixations during recall as indicator of the presence in visual memory, and by averaging over multiple observer as a measure for their importance.  

The fixations during recalling an image while looking at nothing lack a reference frame and cannot be used directly to identify important elements in a visual stimulus. Our idea is to record the exploration of an image using a calibrated eye tracker as well as the immediate recall of the image. The fixations during exploration of the present stimulus can be mapped to locations in the image. And the fixations during recall can be used to validate the fixations during exploration. 

Despite the large body of work on eye motions during recall while looking at nothing, we were unable to find data from experiments. In order to analyze our idea, we have performed an experiment based on standard image data, in which we record eye movements during exploration and recall of those images. 
The complete data set for pairs of exploration and recall sequences will be made publicly available.

\subsection{Eye movements during recall}
The looking at nothing phenomenon, i.e.
the effect of eye movements during the recall of a scene while looking at a neutral visual stimulus, has been known for a long time~\citep{moore1903}, yet was mostly considered an odd coincidence. \cite{Laeng2002} are usually credited with having established that eye movements during recall play a \emph{functional} role in memory retrieval. Their experiment is based on inhibiting eye motion (here, by asking observers to fixate on a point) during either exploration or recall. Inhibiting eye motion during exploration lead to reduced eye motion during recall, and inhibiting eye motion during recall let to decreased recall performance. A variety of other setups have confirmed and refined these findings~\citep{johansson2014,laeng2014,SCHOLZ2015,Bochynska2015,Pathman2015,Scholz2016}. 

In particular, it has been shown that the fixations during recall of the stimulus are connected to the \emph{location} of the objects. \cite{Ferreira2008} noted that observers' gaze would move to the location of previously observed visual elements when cued by keywords, even if the visual element was not present anymore. \cite{de2014} confirmed that inhibiting eye motion during recall decreased memory performance, however, they refined the results and showed that only memory of spatial location was affected, while recollection of visual content was independent of eye movements. \cite{Martarelli2017} showed that gaze still indicates spatial location if observers were asked to imagine a scene with objects being replaced by other similar objects. All these experiments provide evidence that the eye position during recall of a stimulus reveals the location of an object. 

\subsection{Accounting for displacements in recall data}

We want to use the fixations during recall to validate the fixations during exploration. The problem in this task is that the fixations in the recall phase are only qualitatively related to the stimulus -- without the stimulus being present there is no frame of reference. We also find that the motion of the eye during recall contains significant local displacement, i.e.\ the spatial encoding in eye position contains error. We develop an approach for the approximate matching between the fixations in the recall sequence and those in the corresponding exploration sequence. It is based on computing a deformation mapping for the locations of fixations in the recall phase. 
After applying the deformation we retain fixations from the exploration sequence that are close enough to a fixation in the mapped recall.
A threshold on the distance between deformed fixations in the recall and fixations in the exploration sequence allows us to steer the filter. This leads to varying maps of importance, going from considering all fixations in the original exploration sequences to subsets. We explore that resulting spatial importance maps and observe a variety of meaningful effects.

\section{Method}
\label{sec:dataset}

We collect eye movement data during the exploration and recall of images. We first describe how we acquire this data and motivate our design choices for the experiment. To our knowledge, this is the first experiment in this direction. We then provide statistics of the acquired data and show that it is consistent with other data collected for the free-viewing task. 

\subsection{Experiment}

The basic setup for our experiment is to present observers with an image as stimulus and let them recall the image immediately after the presentation of the stimulus. Their head remains in the same position during stimulus presentation and recall. In the following we discuss the details. 

\subsubsection{Participants}
We recruited 30 participants for our experiment (mean age = 26, SD = 4, 9 female). All reported normal or corrected-to-normal vision. Importantly, all participants are naive w.r.t.\ the aim of using the fixations during the recall phase to gauge the importance of fixations during exploration. All studies have been carried out in accordance with the Code of Ethics of the World Medical Association (Declaration of Helsinki). Participants were informed about the procedure before giving their written consent and could stop the experiment at any time. Their time was compensated and all data is used anonymously.

\subsubsection{Apparatus}
The experiment was conducted in a dark and quiet room. A 24-inch display $(0.52 m \times 0.32 m)$ with a resolution of $1920 \times 1200$ pixels was in front of the observer at a distance of $0.7m$. We used an EyeLink1000 desktop mount system (SR Research, Canada) to record the eye movements at a sampling rate of $1000Hz$. A chin and forehead rest was used for stabilization. All experiments were conducted in binocular viewing condition, but only the movements of the dominant eye were recorded. 

\subsubsection{Stimuli}
We use 100 natural images randomly selected from the MIT data set~\citep{judd2012}.
This set includes both indoor and outdoor scenes of various complexity (see Figure~\ref{fig:view_hm} for example images).
These images have been used in eye tracking experiments with a free viewing task and the data is publicly available~\citep{judd2012}. This allows us to compare the eye tracking data we gathered during the exploration phase to existing data. All images were presented at the center of display in their original size with the largest dimension being 1024 pixels. 

\subsubsection{Procedure}

The experiment consists of 100 basic trials, and a single image was presented in each trial. The details of the presentation in one trial are: Prior to the presentation of the image, the screen is black and shows a white dot in the center ($1^\circ$ visual angle). Observers are asked to fixate at the white dot so that the eye motion starts in a consistent way for all images and observers. Then the image is presented for 5 seconds. Observers are asked to explore the image in order to later be able to recognize it (a free viewing task). After presentation of the image white noise is shown for 0.5 seconds to suppress the after image. Then the screen is set to neutral gray for 5 seconds. During this phase, observers are asked to \emph{immediately} recall the image from memory. After that the screen turns black for 1.5 seconds before the procedure is repeated for the next image. Each observer saw all 100 stimuli and the order of them is randomized for each participant.

Instructions as to encourage observers to explore the mental image during the recall phase were given at the beginning of the experiment.  The 100 trials were divided into five blocks. Each block of 20 trials started with a standard 9-point calibration procedure. We repeated the calibration until the average accuracy reported in the following validation was below $0.5^\circ$ and no validation point had an error larger than $1.0^\circ$. After a successful calibration 20 trials were performed. This procedure required roughly 5 minutes.
All participants performed the experiment, however, for two of them we were unable to achieve the desired calibration accuracy (probably due to glasses they had to use). This leaves us with data sets from 28 observers.

An important design choice was the duration of the \emph{exploration} and \emph{recall} periods. In particular the recall phase cannot be too short, because retrieving the image from memory requires time. We decided to follow experiments in cognitive psychology~\citep{laeng2014} and allow 5 seconds for recall. For reasons of temporal equivalence, we also present the stimulus for 5 seconds. The basic task of recalling an image turns out to be significantly more demanding than just exploring the image.
To avoid exhaustion we limited the number of presentations to 100 while still having a sufficient number of repeats for our analysis. 
Participants were allowed to take a break of arbitrary length after each block.

The whole experiment lasted about one hour. At the end of the experiment participants were shown 10 images, half of which were part of the 100 stimuli used in the experiment. Images were presented one after the other in a randomized order, and participants had to decide if the images were among the 100 presented to them.
All parts of the experiment were explained in detail at the beginning, including the memory test at the end. 

\subsection{Data processing and analysis}

We congregated and analyzed the data from 28 observers for the exploration and recall phase of each image.

\subsubsection{Processing raw data and fixation detection}

\begin{figure*}[!t]
\begin{center}
   \includegraphics[width=\linewidth]{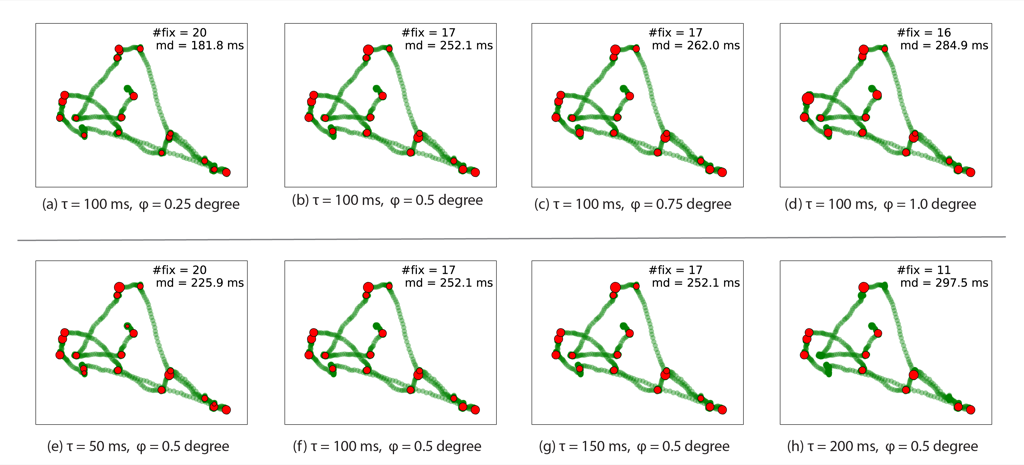}
\end{center}
  \caption{Fixation classification results for different parameter settings. The minimum duration threshold $\tau$ ranges from 50 ms to 200 ms and the maximum dispersion threshold $\varphi$ ranges from $0.25^\circ$ to $1.0^\circ$. Raw samples are depicted in green and fixations in red whose radius correspond to the durations. Fixation number (\#fix) and mean duration (md) are reported in each plot. As shown in the first row, small dispersion $\varphi$ results in more fixations with shorter durations and a large $\varphi$ tends to fuse fixations together. Second row depicts the influence of $\tau$, the minimum duration threshold. Small $\tau$ leads to over segmentation and large $\tau$ results in much less fixations.}
\label{fig:fix}
\end{figure*}

We base our analysis on fixations. To extract them from the raw eye tracking data, we use a dispersion based algorithm~\citep{holmqvist2011}, which involves setting two thresholds: minimum duration $\tau$ and maximum dispersion $\varphi$. Setting these parameters is a standard procedure for eye-tracking experiments with the stimulus being present, we feel experimentation is warranted for the recall phase. We tested a range of parameter combinations for values discussed in the literature and Figure~\ref{fig:fix} shows the classification results of one sequence under different parameter settings. We found that also during recall the settings were not critical and within the common range detected fixations did not vary much. Thus we fixed the parameters for the following analysis. The minimum duration of fixation is set to be $100$ ms and maximum dispersion is set to be $0.5^\circ$ visual angle.

\begin{figure}[!t]
\begin{center}
   \includegraphics[width=\linewidth]{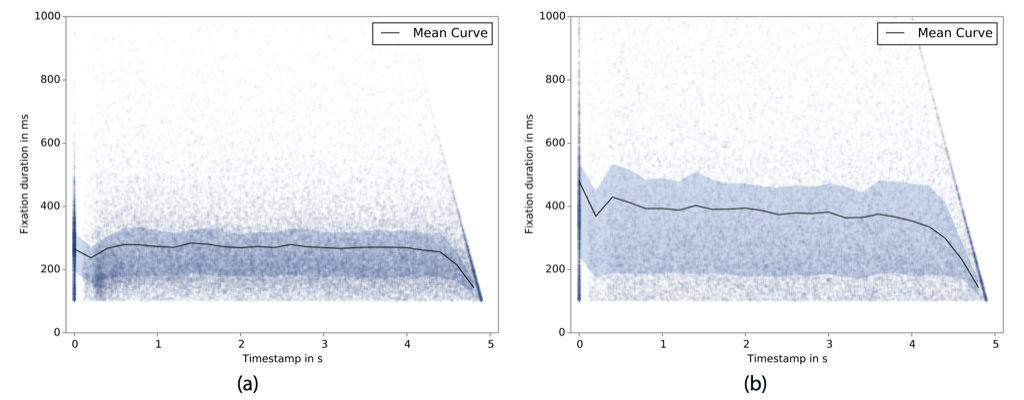}
\end{center}
   \caption{Comparison of fixation durations in exploration (a) and recall (b). Fixation durations are plotted as a function of starting time . All fixations from 28 observers for 100 trials are depicted as crosses. The black curves indicate the mean durations and the center 50 percent intervals are depicted in blue.}
\label{fig:fix_dur}
\end{figure}

We detect fixations in all exploration and recall sequences. During exploration the median and mean number of fixations is $16$ (SD=2.8), and during recall the median and mean number of fixations is $11$ (SD=3.6). The fewer fixations in recall have a correspondingly longer duration (avg = 452.2 ms, SD = 308.0 ms) than fixations in exploration (avg = 278.0 ms, SD=73.4 ms). 
Fixation durations are plotted as a function of the starting time of the fixation during the trial in Figure~\ref{fig:fix_dur}. Exploration phase is separated from the recall phase. 
Each fixation in the dataset from all 28 observers is depicted as a cross and the black curve shows the averaged duration in each phase.

After the first two fixations, there is no significant variation of fixation duration during the trials for both exploration and recall (the durations getting shorter towards the end of the trial is an artifact of stopping data collection after 5 seconds). 
 On average fixations in recall are roughly two times longer than the ones in exploration. 

The fixations from all participants for a single image can be summarized in a spatial histogram, leading to a so-called \emph{heat map} for the image. We do this for the data from our experiment and the publicly available data for the 100 images. Similar to~\cite{judd2012} we remove the first center fixation from each sequence and apply a Gaussian filter with a kernel size equivalent to 1 degree of visual angle.

\subsubsection{Data from exploration}
\begin{figure*}[t]
\begin{center}
   \includegraphics[width=\linewidth]{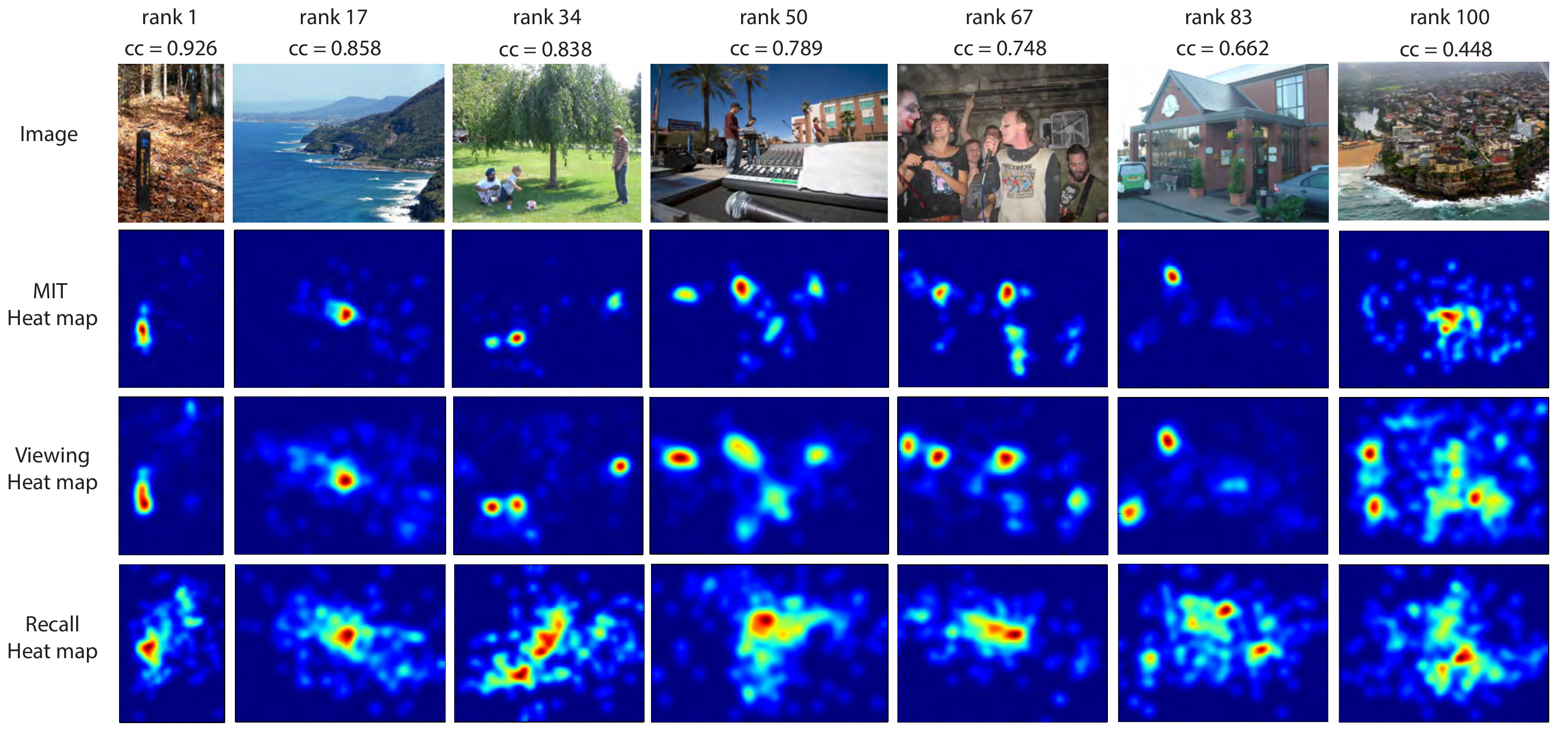}
\end{center}
   \caption{Comparison of heat maps. Images are ranked based on correlation scores between the heat maps generated using the existing data from~\cite{judd2012} and the data we collected during presentation of the images. Seven examples are provided. Each column shows the original image, the MIT heat map, and the heat maps created from the exploration and recall phases of our experiment. }
\label{fig:view_hm}
\end{figure*}

As a sanity check, we compare the heat maps resulting from the publicly available data to the heat maps from the exploration phases in our experiment. To measure the similarity of corresponding maps we use Pearson's correlation coefficient (CC), following~\cite{bylinskii2016}. Averaging over all 100 pairs of saliency maps, we find a mean CC of $0.766$ with standard deviation $0.115$. 
We also perform a qualitative comparison. Figure~\ref{fig:view_hm} shows maps for different ranks among the correlations. Most heat maps are very similar. For those that are dissimilar, we see a stronger center bias in the existing data. We speculate that this is due to the shorter presentation time of 3 seconds vs.\ the 5 seconds in our experiment. We conclude that our results are qualitatively similar to previous experiments with the same data. 

\subsubsection{Data from recall}
Compared to exploration sequences, recall sequences have fewer but longer fixations (see. Figure~\ref{fig:fix_dur})
The last row in Figure~\ref{fig:view_hm} shows the heat maps generated from the recall phases. As can be seen, while they roughly resemble the maps from the exploration phase, they are more biased towards the center and typically fail to \emph{exactly} correspond with features in the image. In particular, they cannot be used directly as heat maps for the images. 

\begin{figure*}[!t]
\begin{center}
   \includegraphics[width=\linewidth]{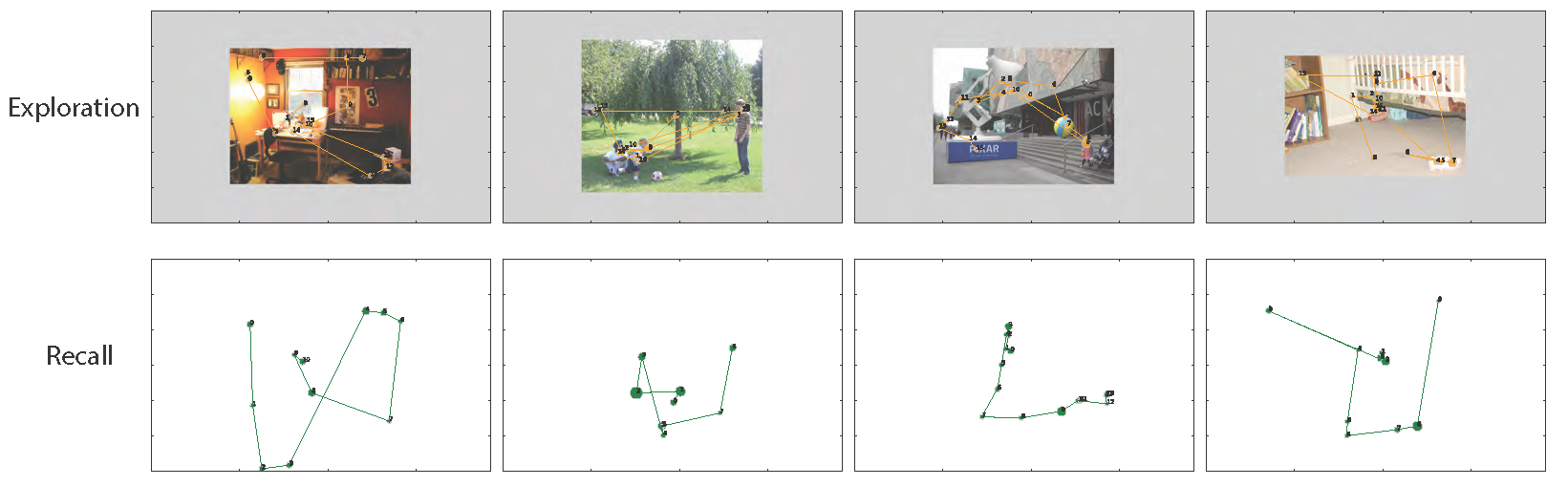}
\end{center}
   \caption{Pairs of exploration and recall fixation sequences from four observers. First row shows fixation sequences collected during exploration and second row shows the corresponding recall sequences. The temporal order of each sequence is indicated by the numbers and consecutive fixations are connected by lines.}
\label{fig:seq_pairs}
\end{figure*}

In some cases, the correspondence between fixations in exploration and recall is clear. Then we found that the temporal order is generally not preserved (see Figure~\ref{fig:seq_pairs} for some examples). This is consistent with previous results in cognitive psychology~\citep{Johansson2012}. 

We also notice that observers tend to 'stall' during recall. This means they stop moving their eyes, leading to fixations that are unlikely corresponding to image content. This sometimes happens for long durations at the center. 
For 5 out of the 28 participants, the number of fixations in recall is less than half compared to exploration. One participant reported after the experiment that he changed his strategy through the experiment and only recalled the single most  interesting element of the image. 

\subsection{Selecting a subset of observers}

\begin{figure*}[!t]
\begin{center}
   \includegraphics[width=\linewidth]{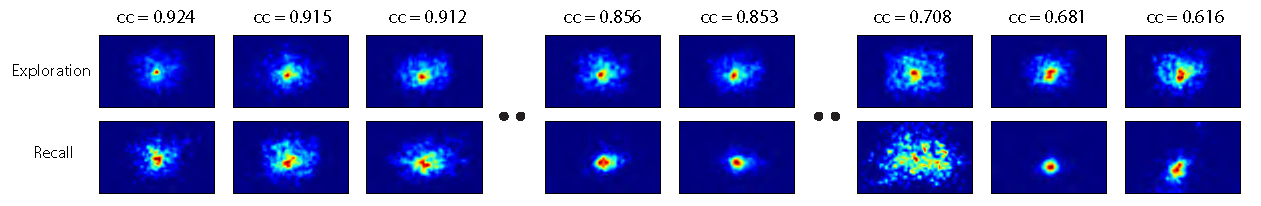}
\end{center}
   \caption{Subset selection criteria. We base our selection on the similarity between the averaged exploration and recall heat maps. Data sets are ranked according to correlation scores. Two different types of inconsistencies are visible in the least similar pairs: the distributions of recall fixations are either peaked at one point or spread more randomly.}
\label{fig:subset}
\end{figure*}

The analysis above suggests that not all observers consistently moved their eyes to revisit important features in the image. Without ground truth it is impossible to decide if the data from a particular observer is useful. In order to avoid basing the analysis on low quality data, we focus our analysis on only half of the observers. 

For selection, we take the fixations for all 100 images from one observer and compare the spatial distribution of fixations during exploration and recall. Figure~\ref{fig:subset} shows heat maps for several observers. The similarity of the spatial distributions varies; some show significantly different patterns (see the three examples on the right hand side in Figure~\ref{fig:subset}). The inconsistencies are of two types:  1) the spatial distribution in recall has smaller variance compared to exploration (indicating observers fixated mostly at one location), 2) the spatial distribution has much higher variance. Both types of differences hamper further using the fixations from recall for matching against the fixations in exploration.

To account for the idiosyncrasies of individual observers we compare the spatial distribution from exploration to the spatial distribution from recall for each observer (rather than judging only the spatial distribution during recall). As an unbiased method to measure their similarity we compute Pearson's correlation coefficient. This results in a ranking of the observers --- the images in Figure~\ref{fig:subset} are sorted based on this ranking. We select the upper half of this ranking for the following processing and analysis. The publicly available data will include all observers. 

\begin{figure}[!t]
\begin{center}
   \includegraphics[width=0.9\linewidth]{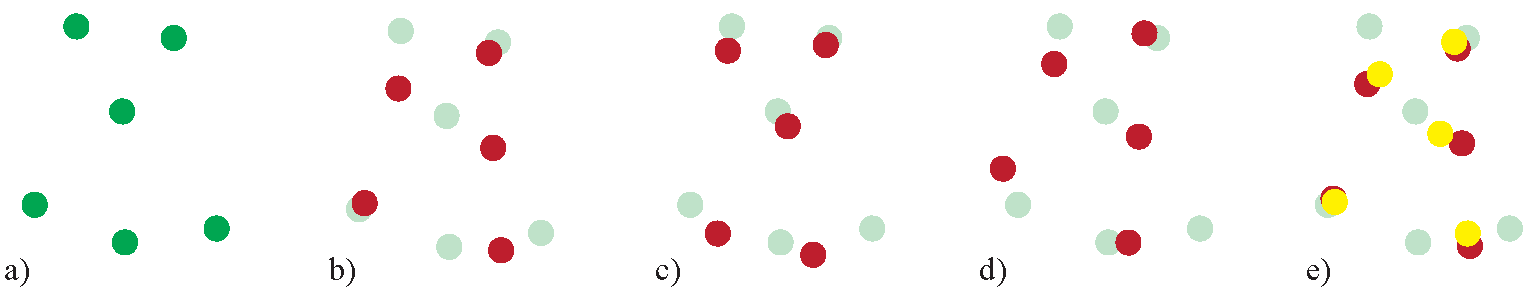}
\end{center}
   \caption{Example of the location mismatch between fixations in exploration and recall. Fixations from exploration are shown in green (a); recall fixations are subset of the fixations during exploration and are shown in red. Their displacement (b) contains global rigid transformation (c) and deformation (d). The deformation mapping is based on the yellow consensus locations (e).}
\label{fig:deformation}
\end{figure}

\section{Filtering the fixations based on recall witnesses}
\label{sec:filter}

Our idea for exploiting the information in the recall is to match the fixations during exploration with the fixations during recall. A fixation in recall matched to a fixation in exploration serves as a \emph{witness}, testifying that the visual element corresponding to the fixation is important. 

\subsection{Deformation mapping}

The problem of establishing the matching is that spatial locations of the fixations in recall are distorted relative to the locations of features in the image. This distortion could be decomposed into a global rigid transformation (due to a lack of reference frame) and local inaccuracies. Figure~\ref{fig:deformation} provides an illustration: The six points in a) depict a sequence of fixations during exploration, of which only five are being recalled. The observed set of fixations during recall projected into the same reference frame is shown in b). We may think of the locations as being composed of the rigid transformation in c) and the deformation in d). The deformation contains no 
global rigid transformation in the sense that moving or rotating the fixations would not make the sum of the squared distances to the matched fixations in exploration any smaller. Note that fixations in recall and exploration that are corresponding cannot be found by simply considering their distances.

When data is corrupted by global transformation and deformation, it is common to approximate the effects by minimizing the squared distances between matched data points. Yet, this is a chicken and egg problem: The estimated mapping is supposed to improve the matching, while the matching is needed to estimate the mapping. The common solution to this dilemma is to start with a first guess about the matching, then compute the deformation, and based on the deformation make a better guess about the matching, and so on until the method converges. A common approach of this type is Iterated Closest Point (ICP)~\citep{besl1992}, which is used to compute a global rigid transformation to match two partially overlapping point sets. 

The approach we suggest for matching the recall fixations to the fixations in exploration is inspired by ICP, yet extends it in two important ways: First, rather than using a closest point matching, it matches fixations in recall to \emph{consensus} locations in the exploration data. Second, rather than computing a global rigid transformation it is based on a deformation mapping that has controllable overall deviation from a global rigid transformation.

The first modification is motivated by the situation in Figure~\ref{fig:deformation} b), namely that matching the recall fixations to the \emph{closest} fixations in exploration may lead to wrong assignments. These wrong assignments will steer the estimated mapping away from the desired solution. Given the data, we believe several situations have to be accommodated:
\begin{enumerate}
\item One fixation in recall maps to exactly one fixation from exploration.
\item One fixation in recall maps to several close fixations from exploration, i.e.\ the observer recalls just one scene element that drew several fixations.
\item A fixation in recall may have no corresponding fixation in exploration (i.e. because the fixation is unrelated to the process of recalling, or the spatial error is too large to be rectified). 
\end{enumerate}
We suggest to accommodate all cases by computing a consensus location for each recall fixation, based on computing a weighted average of the fixation locations in exploration. The weights decay exponentially with distance. This has the effect that if only one fixation in the exploration sequence is close while the others are far away (case 1) the closest location will receive a large weight, while the others relatively small weights, so that the consensus location will be the matching fixation. In case several fixations in exploration are close (case 2), all of them receive equal weights, and the consensus location is the center of these fixations. If no fixation is found for matching (case 3), all fixations receive little weight, and the consensus location is roughly where the recall fixation already is. Figure~\ref{fig:deformation} e) shows the consensus locations for the situation in b). The computation of consensus locations is explained in~\ref{sec:app}. It can be controlled by a parameter $w_p$, measured in visual angle, which could be interpreted as distance of fixations that contribute significantly to the weighted averaging procedure. We set the parameter to $w_p = 2^\circ$ and discuss this choice in Section~\ref{sec:discussion}.

The consensus locations for recall fixations are used to compute a deformation mapping $D: \R^2 \mapsto \R^2$ from the current positions of the recall fixations to the desired ones. Allowing deformation overcomes the problem that the best we could achieve with computing a rigid transformation is to be left with the deformation, i.e.\ the situation depicted in Figure~\ref{fig:deformation} c). However, some global rigidity needs to be preserved, i.e.\ the positions should not deform arbitrarily. This is important, as we would otherwise always match all recall fixations to some fixations in exploration. For example, if we allowed arbitrary scale, it would always be possible to scale the set of recall fixations to a single point and then match it to one of the fixations in exploration. To avoid such degenerated solutions it is important to restrict the mapping to preserve the global structure. The computation of the mapping based on consensus locations and the current positions of the recall fixations is explained in~\ref{sec:app}. The global rigidity of the mapping can be controlled by a parameter $w_d$, measured in visual angle. Roughly speaking, $w_d$ describes the distance of points that may be transformed by two rigid transformations that differ significantly. If $w_d$ is large, fixations are transformed by very similar rigid transformations, restricting the deformation; if it is small, fixations are transformed by independent rigid transformations, allowing deformation. We set the parameter to $w_d = 10^\circ$ and discuss this choice in Section~\ref{sec:discussion}.
 
The necessary steps to compute the deformation mapping are given in pseudocode in~\ref{sec:app}. Once the mapping is computed, we simply use distance as the sole criterion for testimony: a fixation in exploration has a witness, if there is a mapped fixation in recall that is closer than the witness radius $\epsilon$. 

\section{Quantitative results of filtering}

Filtering is supposed to reduce the number of fixations in a meaningful way. It is affected by the choice of witness radius $\epsilon$. Before we analyze specific effects on a qualitative level we present global statistics to show that the effect of filtering is not random.

\begin{figure}[!t]
\begin{center}
   \includegraphics[width=\linewidth]{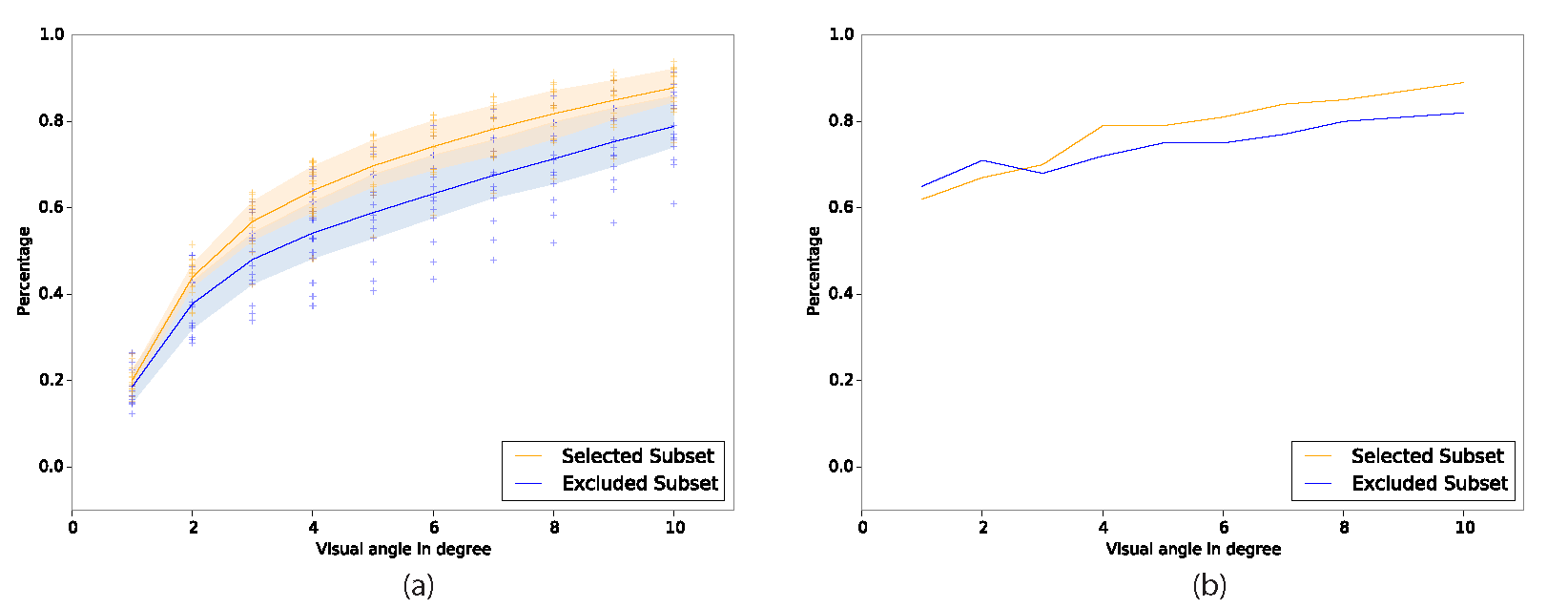}
\end{center}
   \caption{Quantitative results of filtering relative to witness radius $\epsilon \in [1^\circ, 10^\circ]$. Results from sequences in the selected subset of observers are shown in orange and the results from the the excluded subset are shown in blue. (a) Relative number of fixations retained after filtering. Each observer's data is scattered in the graph and the light color marks the center $50\%$ region of the distribution. The effect of the witness radius is uniform across different images and observers. (b) Percentage of all filtered maps whose most salient spots remain within $4^\circ$ after filtering.}
\label{fig:curve}
\end{figure}

One statistic is the number of fixations that remain after filtering. We expect that even for relatively small witness radii a significant number of fixations remain. The left graph in Figure~\ref{fig:curve} depicts the number of retained fixations relative to the witness radius. Here, we differentiate between the subset of observers we selected for further processing vs.\ the subset of excluded observers. Note that the effect of the threshold is quite similar across different images and observers. For values larger than $\epsilon = 10^\circ$ we effectively retain all fixations. For very small values the amount of noise in the recall data leads to results that seem to contain no useful information.
In the subset of observers we consider more reliable fixations remain after filtering. This means that more fixations in the exploration sequences found a closest match in the recall sequences. This is what we expect, as the distributions of the fixations in exploration and recall are similar. 

We also analyze if the sets of fixations change in the sense of their relative importance within an image. For this we identify in each image the region with the most fixations during exploration based on the smoothed spatial histogram. We do the same after filtering and consider that the point has shifted if the difference in spatial location exceeds $4^\circ$. The right graph in Figure~\ref{fig:curve} shows the result relative to the witness radius. We observe that, as expected, for large witness radii there is no significant change, yet for smaller radii the most salient object changes in about $40\%$ of the images. We take this as indication that filtering is not random: on one hand, we expect filtering to affect the relative importance of elements in the scene; on the other hand, in a significant number of images the most fixated object is indeed important.

\begin{figure*}[!t]
\begin{center}
   \includegraphics[width=\linewidth]{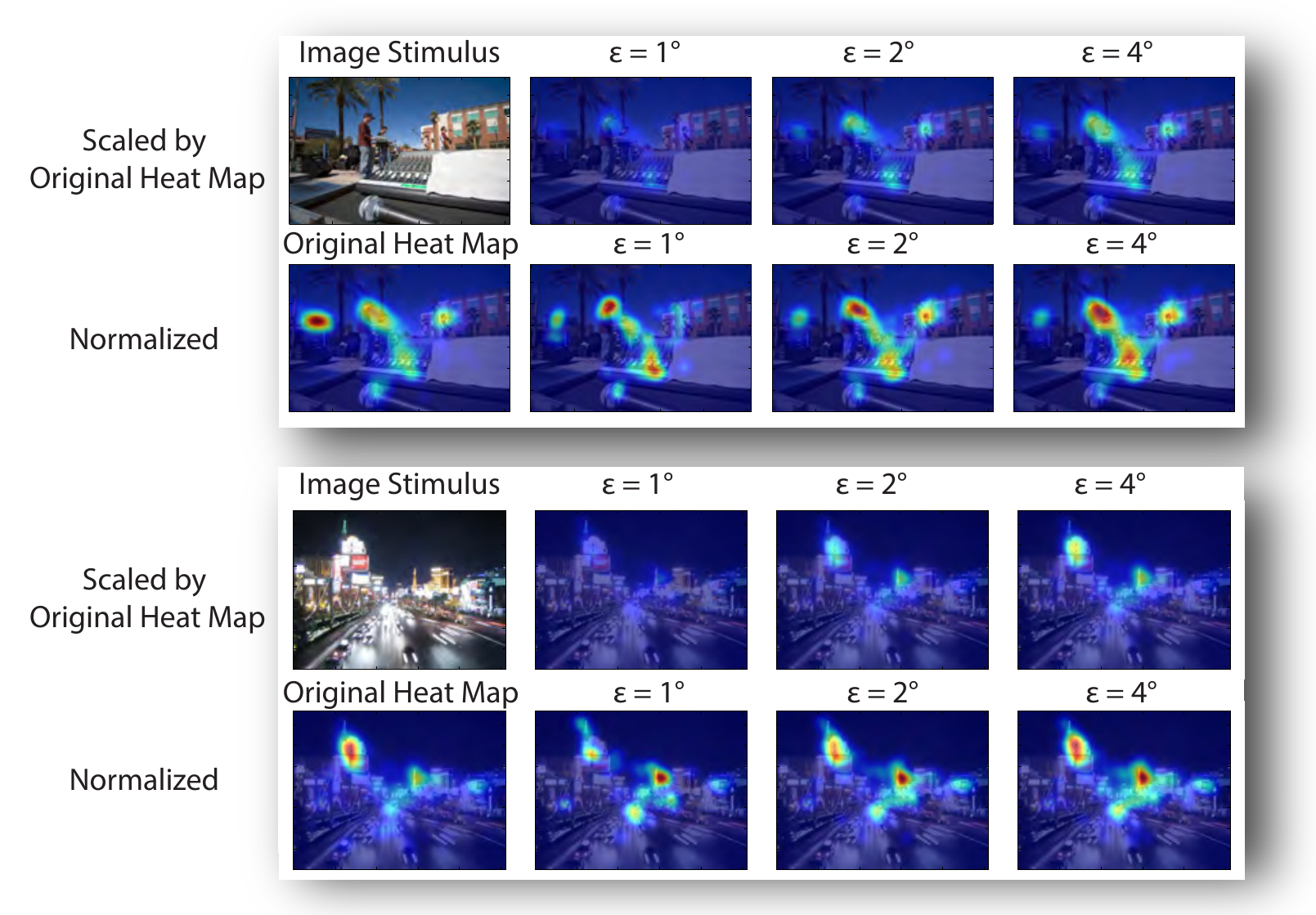}
\end{center}
   \caption{Filtered maps using different thresholds. We show examples of $\epsilon=1^\circ,2^\circ,4^\circ$. Original heat map shows the result based on all fixations during exploration of the image. The color coding of the scaled maps is based on the highest value appearing in all heat maps, which is naturally the one generated without filtering. The normalized map is color coded in its own range.}
  \label{fig:rad_var}
\end{figure*}

Figure~\ref{fig:rad_var} illustrates the effect of varying the threshold on the resulting maps for two of the images using the selected subset. We show heat maps for the values $\epsilon = 1^\circ,2^\circ,4^\circ$ and observe different effects: some regions that attracted a lot of fixations appear to have been deemed relatively unimportant, for example the sign to the left in the top image. Other regions are boosted in importance after filtering such as the objects in the foreground in the top image or the tower to the right of the road in the bottom image. In the following section we attempt a systematic discussion of these effects based on the selected data.

\section{Main results}
\label{sec:results}

We apply the deformation process to each pair of fixation sequences for all images from the subset of selected observers. Then we apply the recall based filtering based on witness radius $\epsilon=1^\circ$. In other words, a fixation is retained only if there is a corresponding fixation in the deformed recall sequence that is within $1^\circ$ visual angle. This is a rather strict setting, removing a lot of fixations. We chose it for achieving a significant effect, at the expense of possibly introducing more noise.

Our analysis is based on comparing the heat maps from the unfiltered and filtered images averaged over the selected 14 observers. For each image, we detect the region showing the largest (absolute) difference in the heat maps (these difference maps are also shown in all illustrations). The inspection of these 100 difference maps showed that the removal of fixations from the exploration sequences can be explained in most cases based on established effects. Before we exemplify these effects, we present an explanation for the removal of a fixation that appears to apply to virtually all cases.

For a region to be removed it necessarily received many fixations during exploration but few during recall. Indeed, most of these regions show complex visual patterns, which explains why they have received a lot of fixations during exploration. Then one of two cases seems to apply: 1) despite the effort, the region could not be decoded into a semantic object and is, consequently, not stored; 2) an object has been recognized, but is deemed unimportant (relative to the rest of the scene, or at least relative to the small amount of time allowed for recalling). In the following we discuss mostly effects of the second type. Clearly, these effects may vary among observers -- in particular, the importance of a recognized object is highly subjective. For effects in the whole dataset of 100 images see the supplementary material.

\subsection{Low-level features}

Low-level features have been shown to contribute to saccade target selection~\citep{badcock1996,Krieger00,kienzle2009}. Consequently, they contribute significantly to fixation-based attentional models. In our model that considers also attention, awareness, and relative importance of the visual object, their strength is no more relevant for the relative importance of a visual element. 

\begin{figure}[!t]
\begin{center}
   \includegraphics[width=0.85\linewidth]{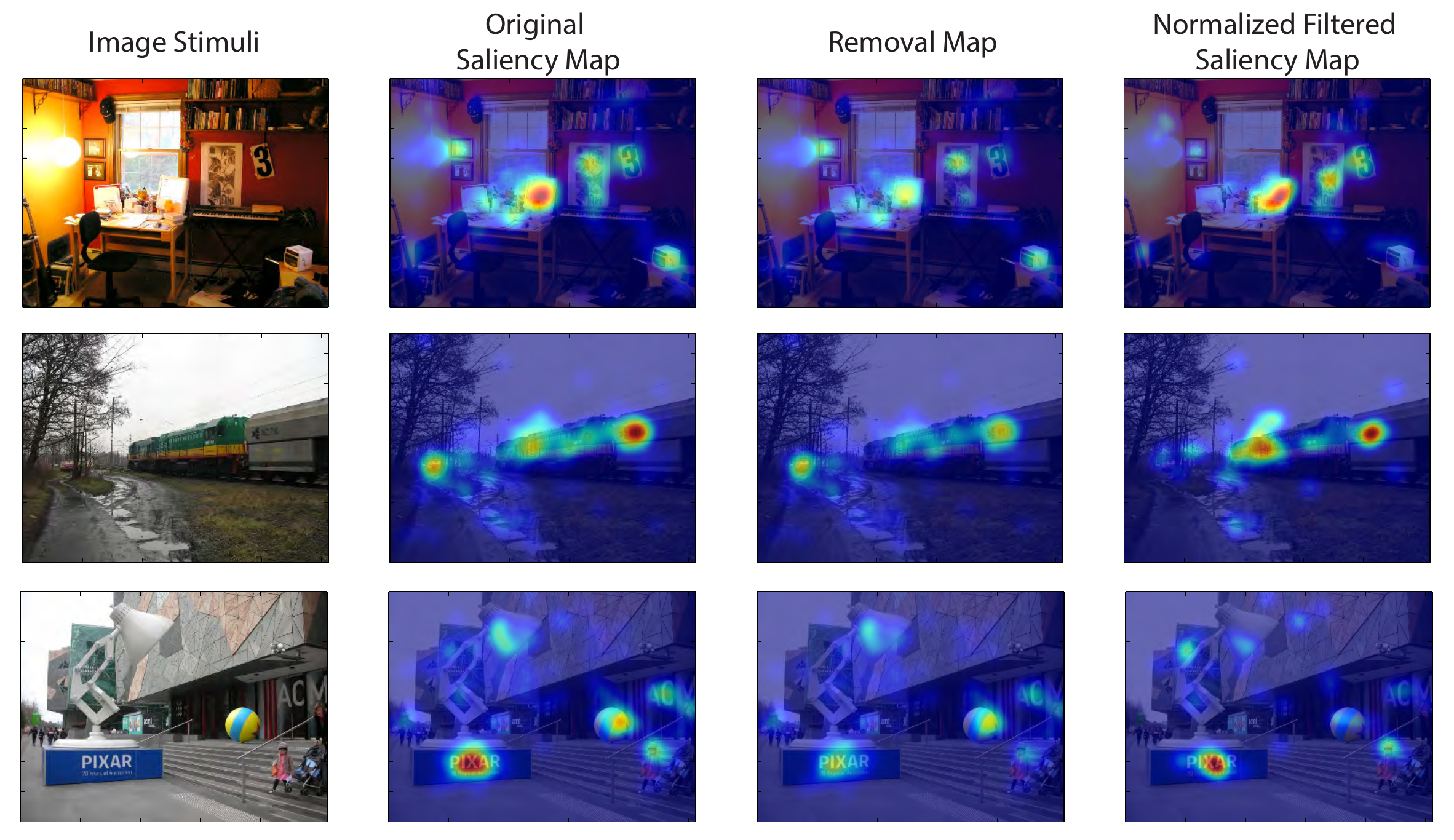}
\end{center}
   \caption{Strong low-level features after filtering. The second column shows the heat map based on all fixations during exploration of the image. Removed fixations are used to generate the removal map (or difference map) and remaining fixations are used to generate the filtered heat map, which is essentially the difference between original heat map and the removal map. The color coding of the removal map is scaled by the highest value in original heat map, while the filtered heat map is normalized in its own range. Fixations triggered by low-level features are largely removed in the filtered heat maps, if the underlying visual object was difficult to understand or turned out to be of little relevance. 
}
\label{fig:low_level}
\end{figure}

Figure~\ref{fig:low_level} shows three examples of such cases. A white box in the right corner over a black background in the first image draws a lot of fixations due to its high contrast. It is less dominant in the filtered heat map as shown in both the third and last columns. We would speculate that observers either were unable to decode the object or decide that a white box-like object is not important for the scene. Similar effects can be observed in the second example on the red car (which is hard to be recognized as such) and in the third example on the ball (which is relatively unimportant compared to its colorfulness and contrast). Instead we see fixations on the train and the shape of the lamp become relatively more important.

\subsection{High-level features}

Among high-level features, we observe noticeable effects on text, signs, and people.

\subsubsection{Text and signs}

\begin{figure}[!t]
\begin{center}
   \includegraphics[width=0.85\linewidth]{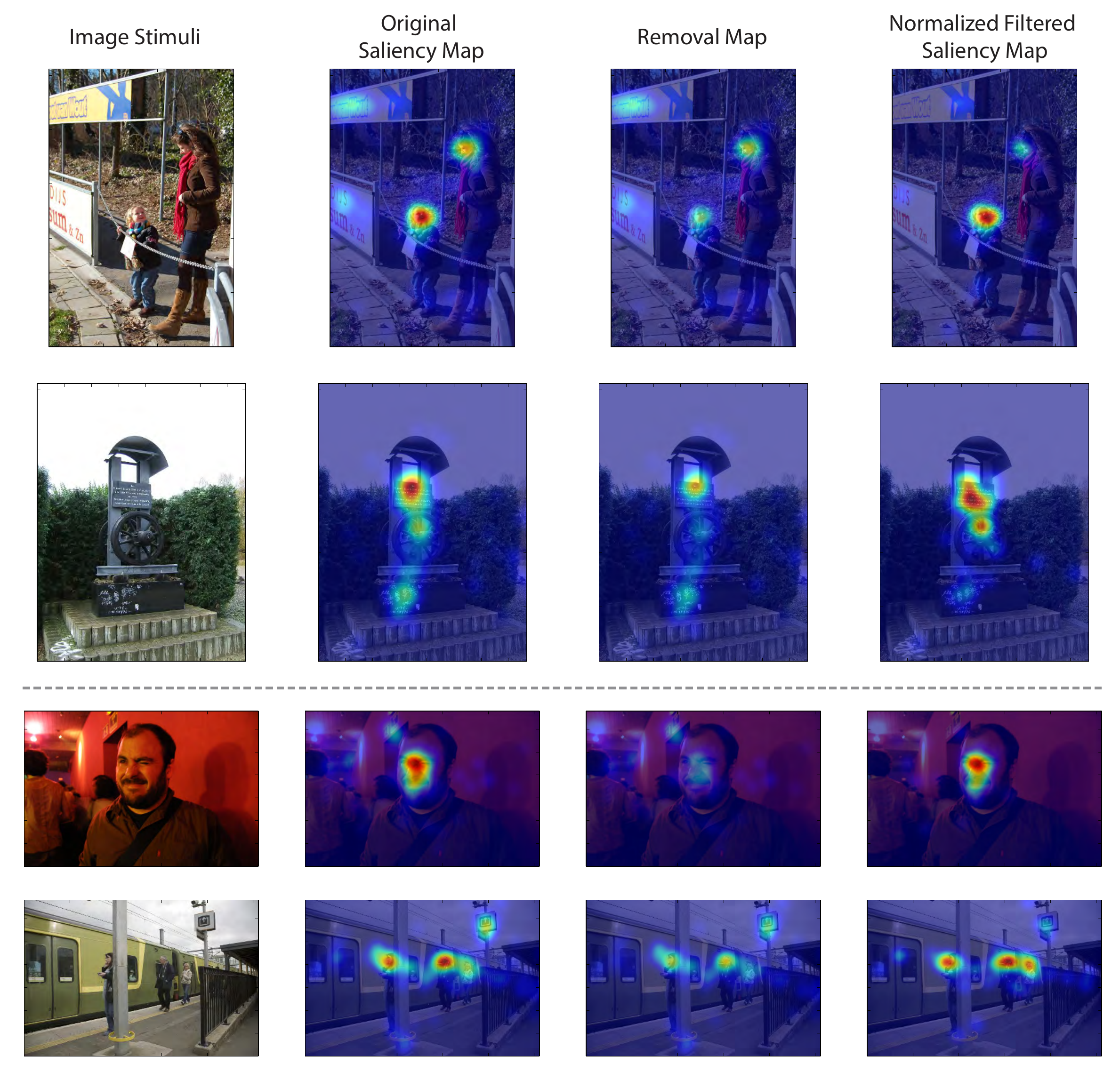}
\end{center}
   \caption{The effect of filtering on text and signs. The meaning of the color coded images is as before. Illegible or unimportant text (upper rows) as well as signs without particular importance for the scene (lower rows) lose importance.}
\label{fig:text}
\end{figure}

Text and signs commonly draw attention as they provide a lot of information in small spatial region. Some signs are purposefully designed to be conspicuous. As shown in Figure~\ref{fig:text}, they have relative high attendance during exploration. Their importance depends significantly on the result of decoding their meaning and then on the experimental conditions, for example the task, but also the experience of the observer. Based on our selection of examples, the elements in Figure~\ref{fig:text} significantly decreased in importance. We believe the text in the first row and the sign in the last row were undecipherable, while in the other two examples the text (whether legible or not) and the exit sign were deemed unimportant for the scene.

\subsubsection{People}
\label{sec:crowd}

\begin{figure}[!t]
\begin{center}
   \includegraphics[width=0.85\linewidth]{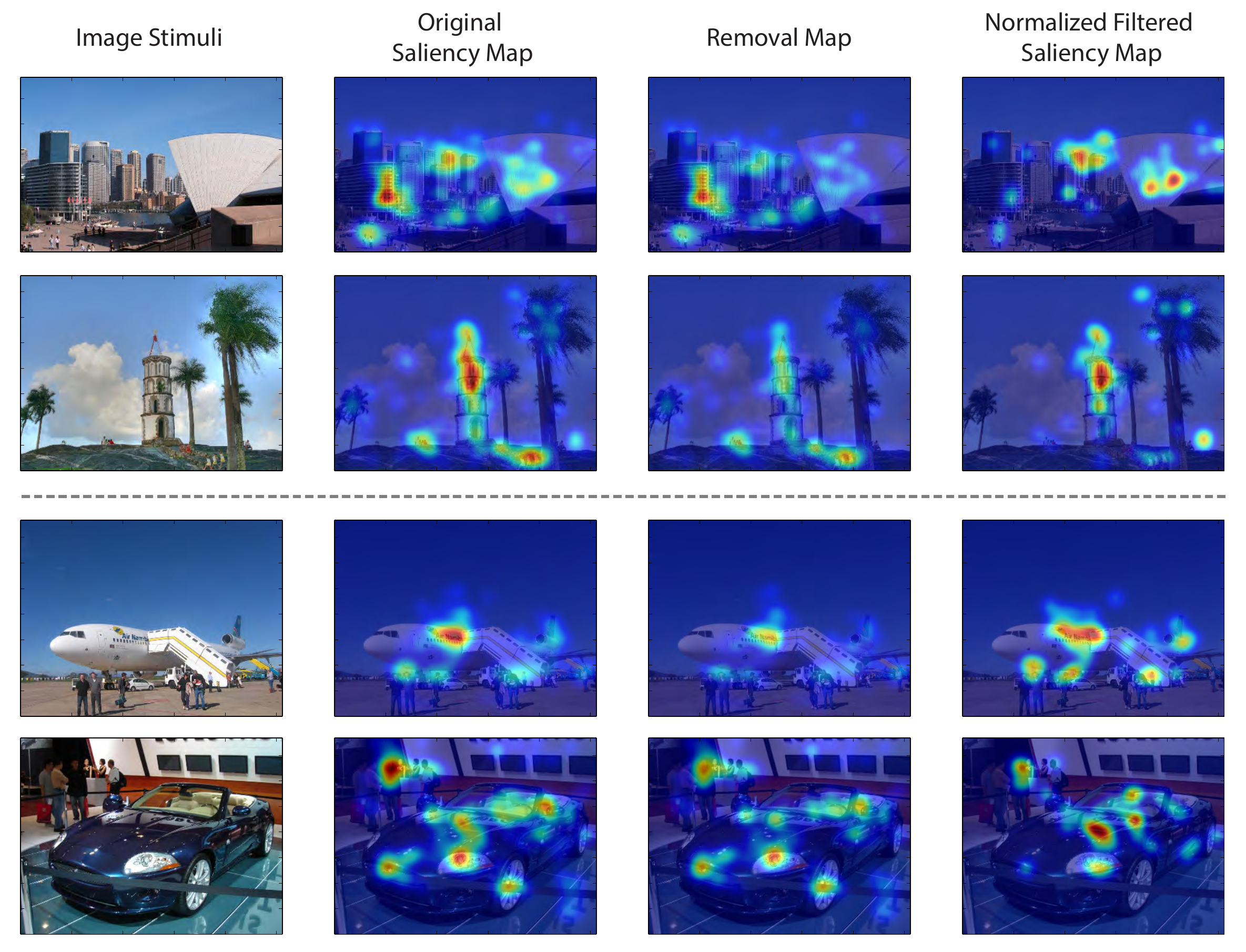}
\end{center}
   \caption{Filtering may or may not affect people. The meaning of the color coded images is as before.
   People in the scene are observed in exploration but their importance in the filtered heat maps depends on the scene composition. In the upper rows, the importance of people decreased, while in the lower rows they remain important.}
\label{fig:crowd}
\end{figure}

As shown in Figure~\ref{fig:crowd}, observers look at humans in images regardless of their size. The relative importance of visual objects recognized as people apparently depends on the scene. In the two examples above the dashed line, people are not a significant element of the scene -- they are expected and not the dominating elements. In the other two examples, humans are contributing more to the composition of the scene, and they remain important after filtering. 

\subsection{Relative importance of similar items}

\begin{figure}[!t]
\begin{center}
   \includegraphics[width=0.85\linewidth]{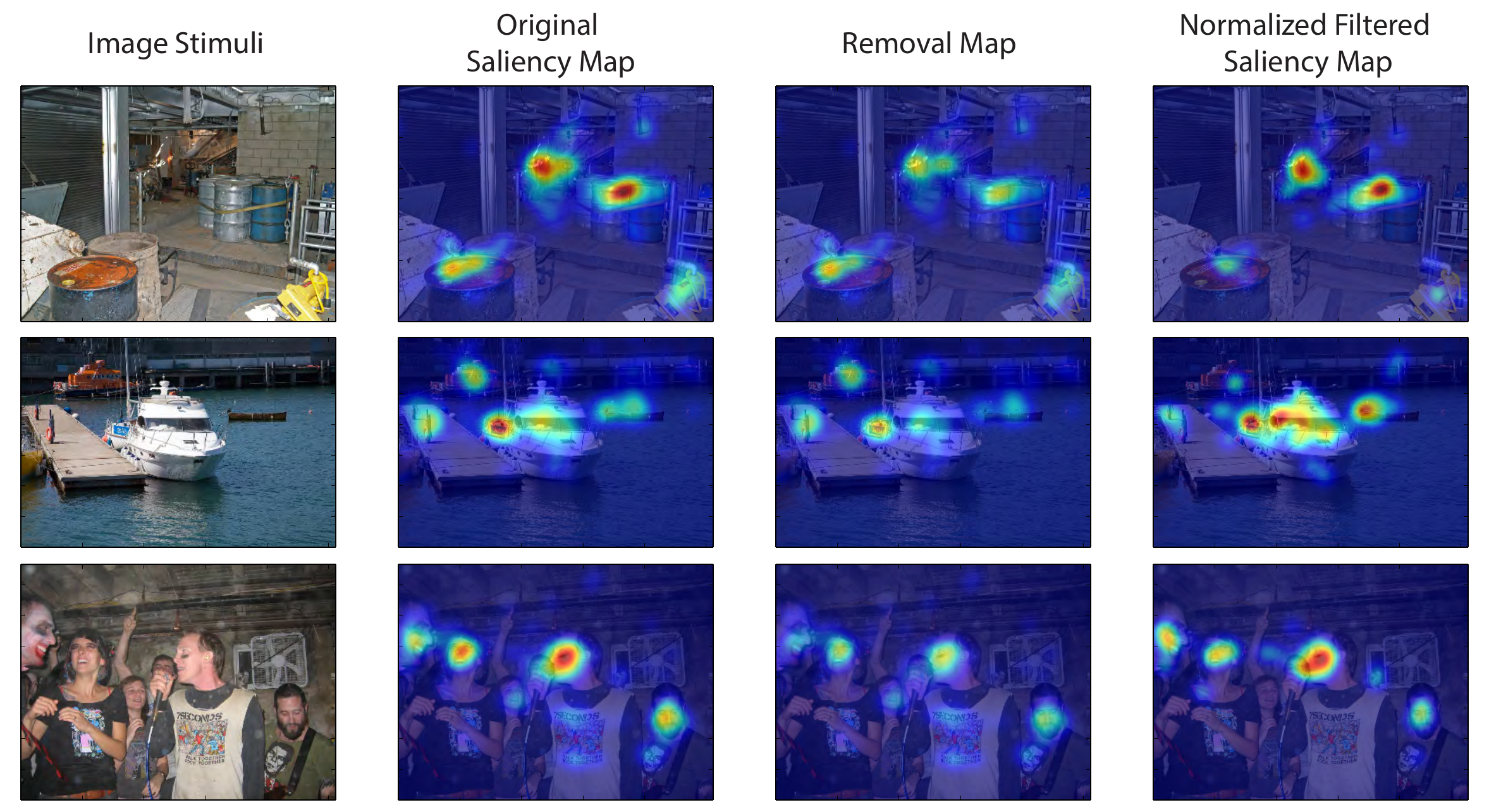}
\end{center}
   \caption{Filtering results for elements with similar meaning. The meaning of the color coded images is as before. In the top row, similar items preserve their similar importance. The smaller boat in the middle row gains in relative importance, while still being dominated by the larger boat. In the last row, the face that stands out among other faces increases in importance.}
\label{fig:repeat}
\end{figure}

Similar objects in a scene may be assigned equal or different importance, independent of their visual representation. We provide several examples in Figure~\ref{fig:repeat}. Similar objects may well be recalled similarly, as is demonstrated in the first example. In the second row, the larger boat dominates the fixations. Based on recall, the smaller boat in the middle row gains in relative importance. Among a set of faces, all of which draw fixations, the one that stands out indeed becomes more important. 

\subsection{Other effects}

\begin{figure}[!t]
\begin{center}
   \includegraphics[width=0.85\linewidth]{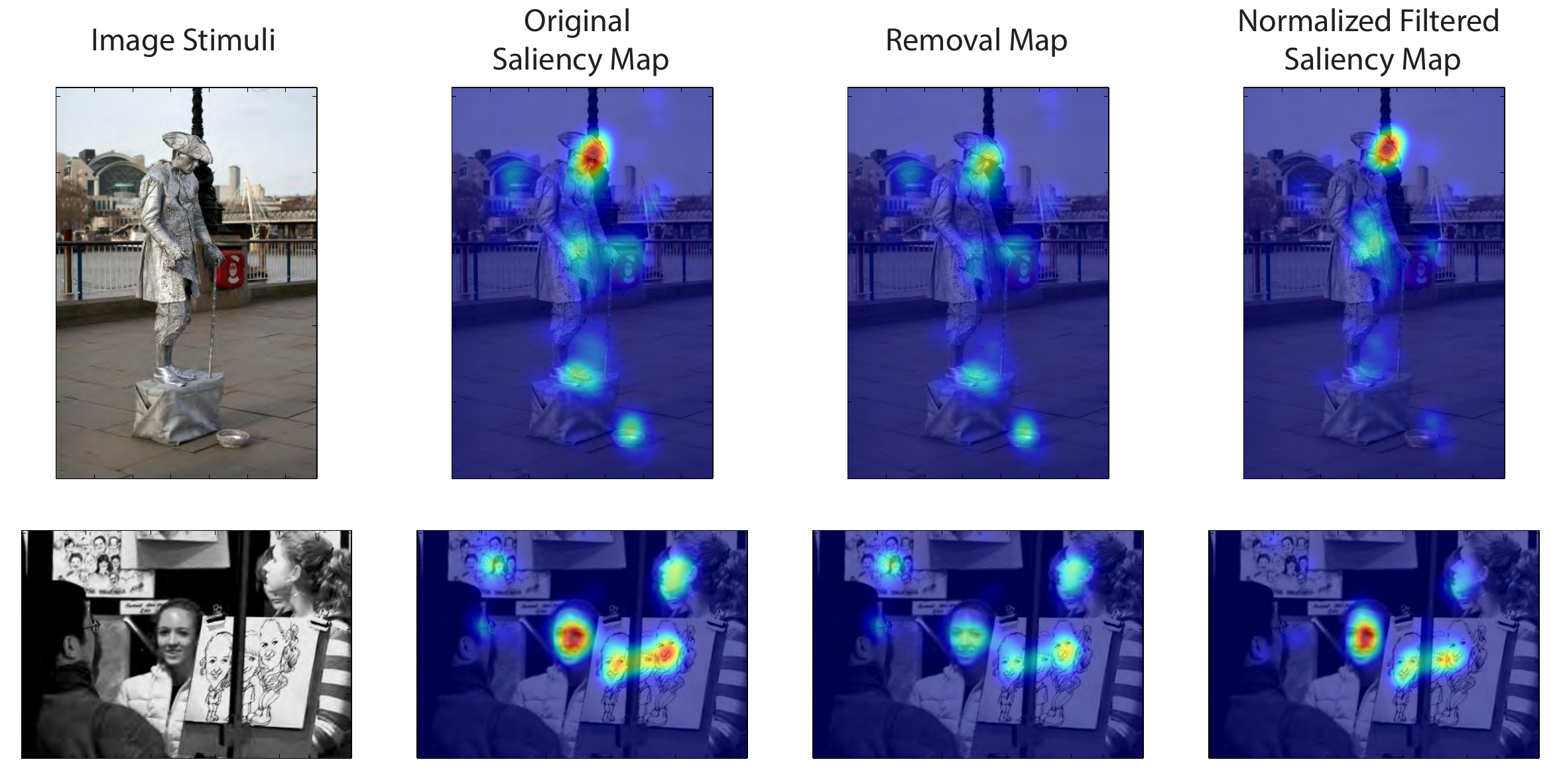}
\end{center}
   \caption{The point of gaze of people in the stimulus is effectively filtered. The meaning of the color coded images is as before. In our recall based attention model, the apparent gaze seems to have less impact and the resulting heat map rather depends on the scene elements being gazed at.}
\label{fig:gazeat}
\end{figure}

When looking at an image we often follow the apparent gaze of people in the scene because it might provide additional information about people's intention or the local environment~\citep{corkum1998,Hutton2011,Gallup2012,bayliss2013}. This effect is notoriously hard to model computationally~\citep{xu2014,recasens2015}. In our model, the object found in these locations determines the importance of the scene element, and not the mere fact that some person in the scene appears to be looking at the region (see~Figure~\ref{fig:gazeat}).

\begin{figure}[!t]
\begin{center}
   \includegraphics[width=0.85\linewidth]{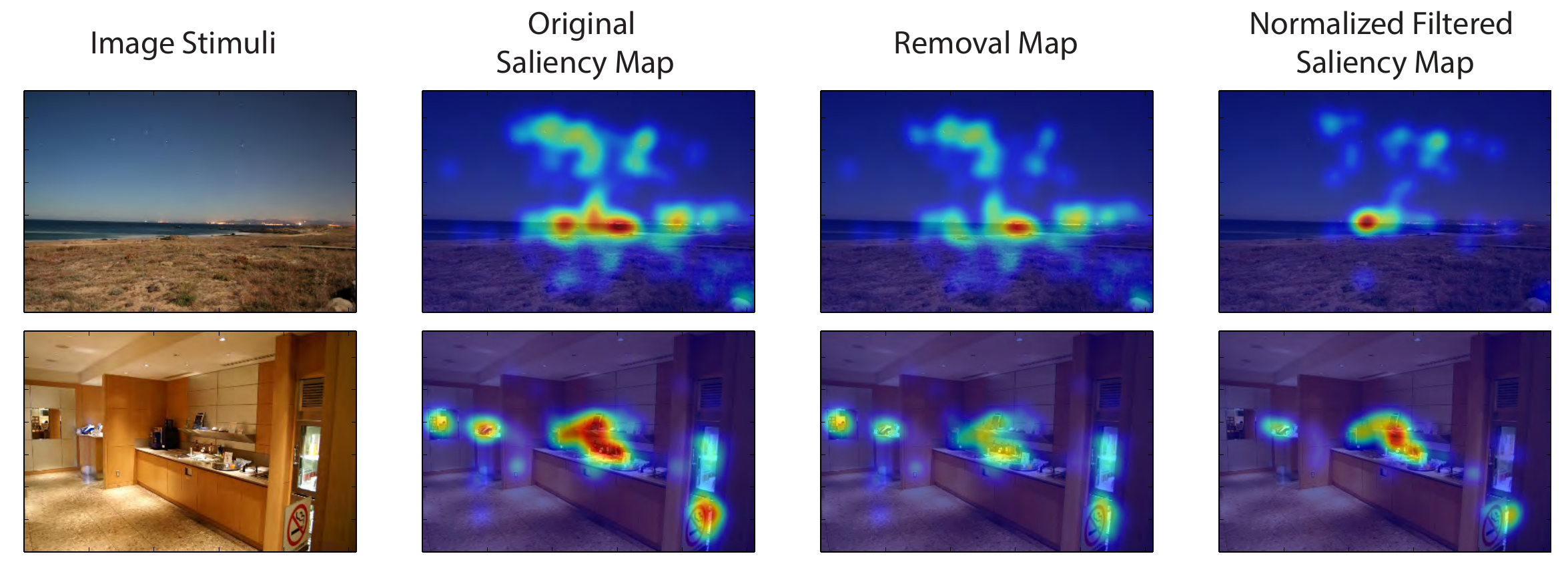}
\end{center}
   \caption{Other effects in the filtered saliency maps. The meaning of the color coded images is as before. Observers scan the horizon, yet only recall elements if they were dominant. Likewise, mirrors are complex visual stimuli, but once they are decoded they become unimportant.}
\label{fig:other}
\end{figure}

Other noteworthy visual elements are horizons and mirrors. Both attract a lot of fixations, however, appear of little importance when the image is being recalled (see Figure~\ref{fig:other}).

\section{Discussion}
\label{sec:discussion}

\subsection{Data collection}
While the fixations we extracted from the collected eye tracking data in the exploration phase are similar to the reference~\citep{judd2012}, there are also some images with considerably different resulting heat maps. Our experimental conditions differ only in the length of exposure, namely 3s vs.\ 5s in our setup. As the data is used for judging the success of computational models for the prediction of saccade targets, we wonder about the variation in the data across observers, and minor variation in experimental conditions, relative to the performance difference of different models. 

We noticed that recall is a strenuous task and not all observers are willing or able to continuously move their eyes. It remains unclear if a lack of motion means that observers paused the recall, however, the fact that the recall performance correlated with motion of the eyes in other experiments~\citep{johansson2014,laeng2014,de2014} suggests that this is the case. It is unclear how to motivate observers to engage more actively in recall. 

Observers only had limited time for recall, thus limiting the number of elements that could potentially being recalled. This situation is similar to the influence of time for recording the fixations during exploration: shorter times leads to fewer fixations and more variation across observers, while longer times increases the amount of noise~\citep{itti2000,Parkhurst2002}. It is not clear how our choice of 5 seconds for recall affected the results.

If it was acceptable to present the stimulus for a significantly longer time, a substantially different protocol for collecting information could be used: observers use standard interaction devices such as computer mice or touch screens to mark interesting points or objects~\citep{jiang2015,kim2015,kim2017}. This higher-level interaction leads to selection based on semantics of features. It has been recently shown to be inconsistent with fixations~\citep{tavakoli2017}. 

\subsection{Subset selection}

We selected only half of the observers based on the correlation of the spatial histograms between exploration and recall. The choice of 14 is arbitrary and it appears that varying it has almost no effect on the outcome. However, we do observe that the data of  few observers is different and would have obscured the results we show. 

For ranking observers we also considered using the pair correlation function~\citep{illian2008,oztireli2012} as it has been recently demonstrated to characterize the spatial distribution of fixations well~\citep{trukenbrod2017}. Yet, we felt there is no obvious way to compare the distributions. Depending on the choice of parameters and method for measuring the similarity the results varied, while largely resembling the order computed with the much simpler approach of taking the correlation. Consequently, we opted for correlation for the subset selection. 

\subsection{Deformation mapping}

As the true matching is unknown, it is impossible to optimize parameters controlling the matching and deformation automatically. We had asked 3 participants in the experiment to manually mark the correspondences among the fixations, in this way generating a reference, yet their data are unreliable (consistent with the literature~\citep{vo2016}).

\begin{figure*}[t]
\begin{center}
   \includegraphics[width=\linewidth]{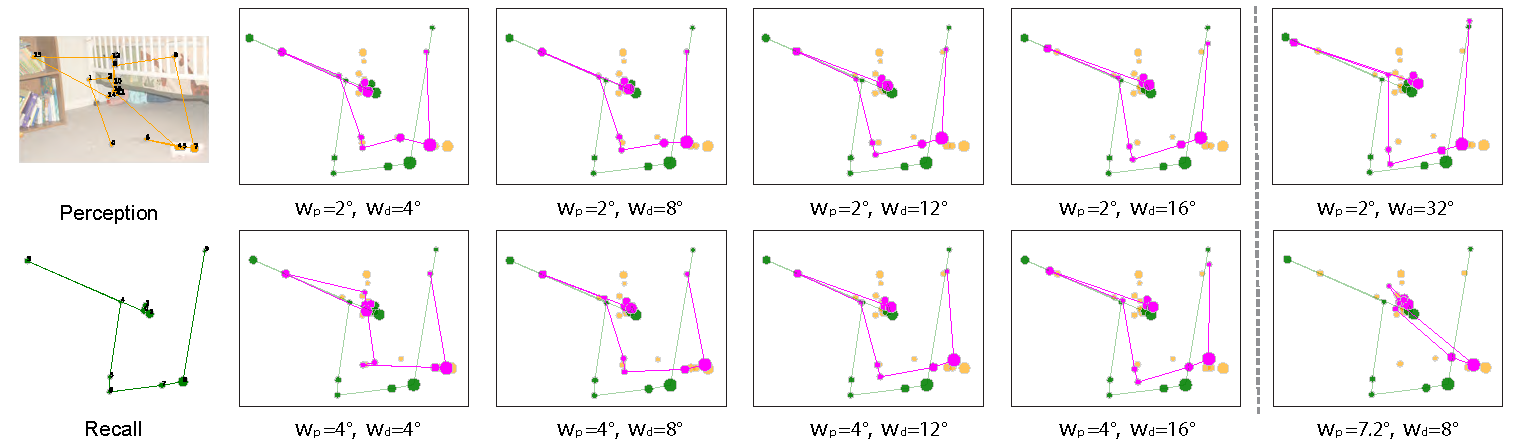}
\end{center}
   \caption{Matching results using different parameters. Original recall fixations (in green) are matched to fixations during exploration (in yellow) and matched fixations are shown in fuchsia. Matching is stable across various parameters. The rightmost column shows cases of undesired deformation for extreme parameter settings. More fixations are considered for matching with a larger $w_p$ while the deformation is more rigid with a larger $w_d$.}
\label{fig:param}
\end{figure*}

\begin{figure}[t]
\begin{center}
   \includegraphics[width=\linewidth]{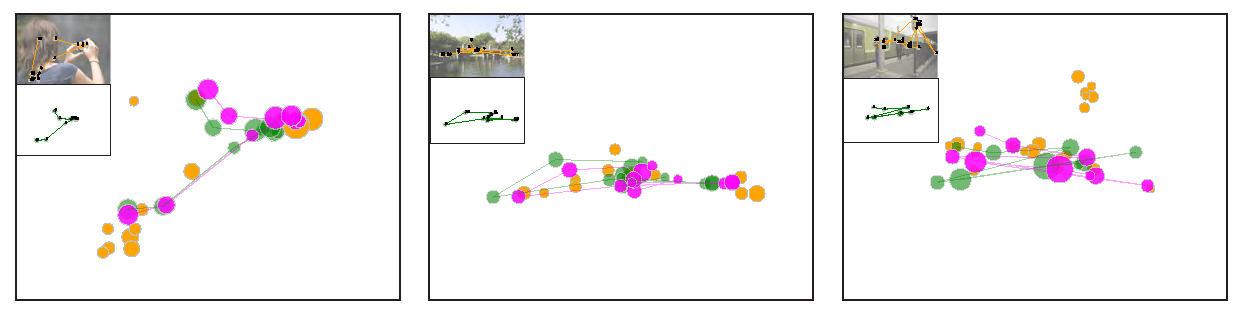}
\end{center}
   \caption{The deformation model performs well on different examples. Parameters $w_p$ is set to $2^\circ$ and $w_d$ to $10^\circ$. Each example shows the fixations during exploration in yellow and recall fixations in green. Each pair overlaid with the stimulus is shown at the top-left corner. Matched fixations are shown in fuchsia.}
\label{fig:match_res}
\end{figure}

On the other hand, the results of the deformation process are stable across a wide range of parameters (see Figure~\ref{fig:param}). We found that the radius for matching recall fixations to exploration fixations $w_p$ should be chosen in the range of $2^\circ-4^\circ$ visual angle. This means we expect that matching fixations are usually not separated by more than twice this amount. To limit the deformation of the mapping we have tried values of $w_d \in [4^\circ,16^\circ]$. Based on experimentation we have settled for $w_p = 2^\circ$ and $w_d = 10^\circ$. Figure~\ref{fig:match_res} shows the results of three examples using this parameter setting. 

\begin{figure}[t]
\begin{center}
   \includegraphics[width=\linewidth]{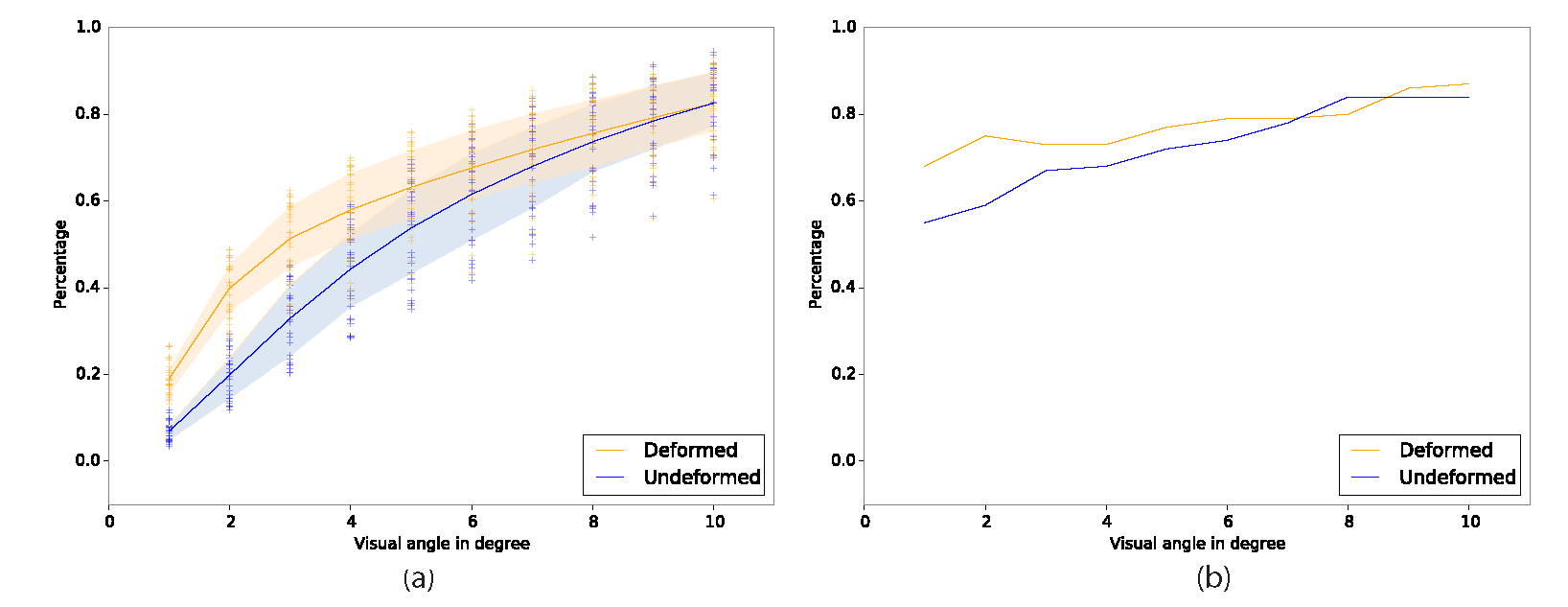}
\end{center}
   \caption{Results of filtering relative to witness radius $\epsilon \in [1^\circ, 10^\circ]$ based on original recall fixations are shown in blue. No deformation mapping is applied. Results of filtering based on deformed recall fixations are depicted in orange for comparison. (a) Relative number of fixations retained after filtering. Each observer's data is scattered in the graph and the light color marks the center $50\%$ region of the distribution. Much less fixations are retained when $\epsilon$ is small. (b) Percentage of all stimuli maps whose most salient spots remain within $4^\circ$ after filtering. More images contain changes of importance without applying deformation. Small $\epsilon$ in particular highlights the effects of deformation mapping.}
\label{fig:no_matching}
\end{figure}

In order to understand if deformation was really necessary to achieve the results we have presented, we perform the filtering based on the original, undeformed recall sequences. Similar to Figure~\ref{fig:curve}, Figure~\ref{fig:no_matching} plots the percentage of remaining fixations as a function of the witness radius $\epsilon$ varying from $1^\circ$ to $10^\circ$ visual angle, as well as the relative changes of importance of fixations. 
Comparing the results 
we notice that without applying deformation, 
much less fixations remain after filtering. For $\epsilon=1^\circ$ nearly $20\%$ of fixations during exploration remain after applying deformation, while only $5\%$ have matched fixations in the original recall. Similar effects can be observed on the changes of the most important visual element: filtering based on the deformed recall sequences lead to significantly more of the images being unaffected. We speculate that without deformation several important elements remain unmatched, causing significant but unwarranted changes to the map. 

\subsection{Visual attention and visual working memory}

Visual attention and visual working memory are entangled: attention controls encoding in working memory by selecting what is the relevant information, and visual working memory biases the focus of attention~\citep{Downing2000,Schmidt2002,olivers2006,Hollingworth20132}.
Several studies suggest that visual attention and visual working memory share the same representations~\citep{farah1985,harrison2009,Hollingworth2013,ALBERS2013}, and it has been debated whether they should be regarded as one cognitive function since they also share some of the same mechanisms (see~\cite{olivers2008} for a review).

Our results suggest that the prioritization in visual attention differs from the prioritization in working memory: the spatial importance maps are affected by filtering, and filtering is based on a functional connection to the spatial memory of visual elements. While more experiments are needed to understand this difference, we hope the experimental setup based on immediate recall of previously presented stimuli provides an additional useful tool.

\section{Conclusions \& Outlook}
\label{sec:conclusion}

The 'looking-at-nothing' phenomenon is an interesting way to test the awareness of a feature in an image previously fixated during exploration. This gives rise to a new way of measuring the importance of visual elements as assigned by a human observer. We have generated a data set based on eye tracking experiment that allows studying this idea and other researchers can use it as a basis for additional analysis.

Overall, this importance model appears to conform to the idea of decomposing the image into relevant pieces --- low-level features are only important for areas to be visited during exploration. We 
argue that classical saliency maps overemphasize the role of fixations. And concluding that objects which drew a lot attention based on fixations are indeed \emph{recognized} by the observer may be wrong.

The main challenge in this approach is the large amount of distortion, noise, and artifacts in the recall data. There seems to be two ways to address this problem: better processing of the data hopefully resulting in better matching; and better data collection.

It might be possible to exploit the raw data to better match the recall to exploration sequence, as apparently not just the fixations correspond but also partial scan paths~\citep{Johansson06}. The matching might be improved using machine learning techniques if we had ground truth data. It remains unclear how to generate such data. As mentioned, we have tried to get this information from the observers, but this attempt has been unsuccessful. 

We believe the experimental setup leaves little room for improvement in order to get cleaner data. The main problem is to motivate participants to equalize recall and eye movements without making the experiment significantly more complex. We made one observation that would be worth exploiting: It appears that the recall performance improves in repeated trials. Similar effects have been observed in other recall experiments~\cite{mantyla2006,kaspar2011}. A good protocol for such an experiment would still have to be designed. 

The setup clearly exploits short term memory. The results for long-timer recall, specifically if locations of fixations still encode the locations of features in images, are inconsistent~\citep{martarelli2013,laeng2014,wantz2016}. We expect the amount of noise in the data to increase and, therefore, finding the matching between recall and exploration to be even more difficult. Nonetheless, a setup similar to ours, yet with considerably delayed recall might be worthwhile. Such an experiment would be related to image memorability~\citep{isola2014,khosla2015,bylinskii2015}, which explores the memorability of visual objects, and the results would likely become even more dependent on personal factors. 

The most successful techniques for predicting fixations to date are based on neural networks~\citep{huang2015, kruthiventi2015}. We considered using similar approaches to predict the filtered heat maps. Better prediction would indicate that our results are more consistent. However, given the amount of noise in the data we use for filtering and the high-level effects we found in our analysis, we are unsure about the meaning of this analysis. We still think it would be generally fruitful to use filtered heat maps for learning important features that are closer to what humans really encode. 

\section*{Acknowledgments}
We would like to thank Marianne Maertens for valuable advice. Furthermore, we thank David Lindlbauer for help performing the experiment. This work has been partially supported by the ERC through grant ERC-2010-StG 259550 (XSHAPE).

\begin{appendix}
\section{Computing the deformation mapping}
\label{sec:app}

Let $\mv{p}_i \in \R^2$ be the positions of the fixations in the exploration sequence and $\mv{r}_j \in \R^2$ the positions of fixations in the recall sequence (in a common coordinate system). We wish to compute a deformation $D: \R^2 \mapsto \R^2$ that is applied to the recall locations $\mv{r}_j$ with the aim to align the data with the fixations positions during exploration.

We need two ingredients for computing the deformation: partial matching and deformation mapping.
For the deformation, we suggest Moving Least Squares (MLS)~\citep{Levin:1998:APM}. Here, we use this framework applied to rigid transformations, i.e.\ local rigid transformations are fitted using weighted least squares. This approach has become popular in geometric modeling where it is usually derived as minimizing the deviation of the mapping from being locally isometric~\citep{schaefer2006,sorkine2007,Chao:2010:SGM}. 

We model the deformation $D$ as a rigid transformation that varies smoothly over space:
\begin{equation}
D(\mv{x}) = \mv{R}_{\mv{x}} \mv{x} + \mv{t}_{\mv{x}}.
\label{eq:transformation}
\end{equation}
The subscript $\mv{x}$ indicates that rotation and translation vary (smoothly) with the location $\mv{x}$ in the plane. They are computed by solving a weighted special orthogonal Procrustes problem~\citep{gower2004procrustes}, where the weights depend on the distance of the points to $\mv{x}$. Assume the desired position for $\mv{r}_j$ is the position $\mv{q}_j$, then $\mv{R}_{\mv{x}},\mv{t}_{\mv{x}}$ are computed by solving
\begin{equation}
\argmin_{\mv{R_{\mv{x}}}\tp\mv{R_{\mv{x}}} = \mv{I}, \mv{t}_{\mv{x}}}
\sum_j \theta(\| \mv{x} - \mv{r}_j\|) \| \mv{R}_{\mv{x}}\mv{r}_j + \mv{t}_{\mv{x}} - \mv{q}_j \|_2^2.
\label{eq:min}
\end{equation}
Here, the weight function $\theta$ should be smoothly decaying with increasing distance. We use the common choice
\begin{equation}
\theta_d(x) = e^{-\frac{x^2}{w_d^2}},
\label{eq:weight}
\end{equation}
which gives us control over the amount of deformation in the mapping with the parameter $w_d$. 
The minimization can be solved directly using the singular value decomposition (SVD), see ~\cite{SorkineRabinovich:SVD-rotations:2016} for an accessible derivation.

Note that for computing the mapping we simply assumed the desired positions $\mv{q}_j$ were given.
We compute them as the distance weighted centroid of exploration fixations:
\begin{equation}
\mv{q}_j = \frac{\sum_i \theta_p(\| D(\mv{r}_j) - \mv{p}_i \|) \mv{p}_i}{\sum_i \theta_p(\| D(\mv{r}_j) - \mv{p}_i \|)},
\label{eq:neighbor}
\end{equation}
where $\theta_p(d)$ quickly decreases such that points further away are receiving relatively insignificant contribution.

\begin{figure}[t]
\begin{center}
   \includegraphics[width=\linewidth]{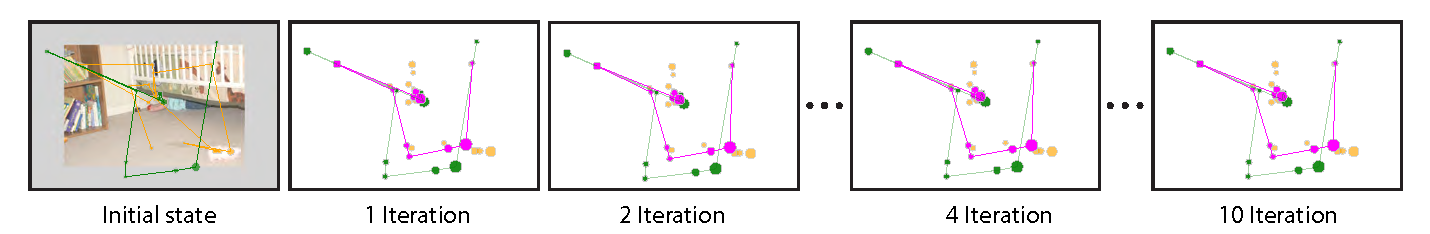}
\end{center}
   \caption{Iterations of the deformation. The initial state shows two fixation sets of viewing (orange) and recall (green). Radius of each circle correlates to the fixation duration. The set of fuchsia circles are the deformed recall fixations in each iteration. Note that the matching converges quickly after a few iterations.}
\label{fig:conv}
\end{figure}

Note that we are considering the distances of the fixations $\mv{p}_i$ to the \emph{deformed} locations of the fixations in the recall sequence. This means that setting the target locations depends on the deformation mapping and computing the deformation mapping depends on the target locations. Consequently, we alternate the two steps as shown in Algorithm~\ref{alg:icp}. 
We start this process with $D$ being the identity. Then we compute the desired positions $\mv{q}_j$ as explained above. The procedure converges after very few iterations (see Figure~\ref{fig:conv}). 
Note that the deformation $D$ needs to be evaluated only in the location $\mv{r}_j$. This means for the next step we only need to compute $\mv{R}_{\mv{x}}$ and $\mv{t}_{\mv{x}}$ for $\mv{x} = \mv{r}_j$.

\begin{algorithm}
\KwData{Fixations locations in exploration $\mv{p}_1,\ldots,\mv{p}_n$ and recall $\mv{r}_1,\ldots,\mv{r}_m$, convergence criteria $\lambda$}
\KwResult{Locations of mapped recall fixations $\mv{r}'_1,\ldots,\mv{r}'_m$}
\SetKwFunction{SVD}{SVD}
\DontPrintSemicolon
$\forall (i,j) \in [1,m]^2: w_{ij} \leftarrow \exp(-\| \mv{r}_i - \mv{r}_j \|^2/w_d^2) $\;
\For{$j \in [1,m]$}
{
	$\mv{r}'_j \leftarrow \mv{r}_j$\;
	$\bar{\mv{r}}_i \leftarrow \left( \sum_{j=1}^m w_{ij} \mv{r}_j \right) / \left( \sum_{j=1}^m w_{ij} \right)$\;
	$\mv{R}_i = \left( w_{i1}(\mv{r}_1 - \bar{\mv{r}}_i), \ldots, w_{im}(\mv{r}_m - \bar{\mv{r}}_i) \right)$\;
}
\Repeat{ $\sigma < \lambda$ }
{
	\For{$i \in [1,m]$} 
	{
		$\mv{q}_i \leftarrow
		 \frac{\sum_{j=1}^n \exp(-\| \mv{r}'_i - \mv{p}_j \|^2/w_p^2) \mv{p}_i}
		 {\sum_{j=1}^n \exp(-\| \mv{r}'_i - \mv{p}_j \|^2/w_p^2)}$\;
	}
	$\sigma \leftarrow 0$\;
	\For{$i \in [1,m]$} 
	{
		$\bar{\mv{q}} \leftarrow \left(\sum_{j=1}^m w_{ij} \mv{q}_j \right) / \left( \sum_{j=1}^m w_{ij} \right) $\;
		$\mv{Q} = \left( \mv{q}_1 - \bar{\mv{q}}, \ldots, \mv{q}_m - \bar{\mv{q}} \right)$\;
		$\mv{U}\mv{\Sigma}\mv{V}\tp \leftarrow \SVD \left(\mv{R}_i \mv{Q}\tp\right)$\;
		$\mv{s} \leftarrow \mv{V}\mv{U}\tp (\mv{r}_i - \bar{\mv{r}}_i) + \bar{\mv{q}}$\;
		$\sigma \leftarrow \sigma + \| \mv{r}'_i - \mv{s} \|$\;
		$\mv{r}'_i \leftarrow \mv{s}$\;
	}
}

\caption{Deformation mapping}
\label{alg:icp}
\end{algorithm}

\end{appendix}

\section*{References}


\end{document}